\documentclass[aps,prl,twocolumn,showpacs,superscriptaddress,amsmath,amssymb]{revtex4}

\usepackage{graphicx}
\begin{document}

\title{Bulk electronic structure of optimally doped Ba(Fe$_{1-x}$Co$_{x}$)$_2$As$_2$}

\author{C. Utfeld}
\affiliation{H.~H.~Wills Physics Laboratory, University of Bristol, Tyndall Avenue, Bristol BS8 1TL, United Kingdom}
\author{J. Laverock}
\affiliation{H.~H.~Wills Physics Laboratory, University of Bristol, Tyndall Avenue, Bristol BS8 1TL, United Kingdom}
\author{T.~D. Haynes}
\affiliation{H.~H.~Wills Physics Laboratory, University of Bristol, Tyndall Avenue, Bristol BS8 1TL, United 
Kingdom}
\author{S.~B. Dugdale}
\affiliation{H.~H.~Wills Physics Laboratory, University of Bristol, Tyndall Avenue, Bristol BS8 1TL, United Kingdom}
\author{J.~A. Duffy}
\affiliation{Department of Physics, University of Warwick, Coventry CV47AL, United Kingdom}
\author{M.~W. Butchers}
\affiliation{Department of Physics, University of Warwick, Coventry CV47AL, United Kingdom}
\author{J.~W. Taylor} 
\affiliation{ISIS Facility, Rutherford Appleton Laboratory, Chilton, Oxfordshire
OX11 0QX, United Kingdom} 
\author{S.~R. Giblin}
\affiliation{ISIS Facility, Rutherford Appleton Laboratory, Chilton, Oxfordshire 
OX11 0QX, United Kingdom}
\author{J.~G. Analytis}
\affiliation{Stanford Institute for Materials and Energy Sciences, SLAC National Accelerator Laboratory, 2575 Sand Hill Road, Menlo Park, CA 94025}
\affiliation{Geballe Laboratory for Advanced Materials and Department of Applied
Physics, Stanford University, CA 94305}
\author{J.-H. Chu}
\affiliation{Stanford Institute for Materials and Energy Sciences, SLAC National Accelerator Laboratory, 2575 Sand Hill Road, Menlo Park, CA 94025}
\affiliation{Geballe Laboratory for Advanced Materials and Department of Applied
Physics, Stanford University, CA 94305}
\author{I.~R. Fisher}
\affiliation{Stanford Institute for Materials and Energy Sciences, SLAC National Accelerator Laboratory, 2575 Sand Hill Road, Menlo Park, CA 94025}
\affiliation{Geballe Laboratory for Advanced Materials and Department of Applied
Physics, Stanford University, CA 94305}
\author{M. Itou}
\affiliation{Japan Synchrotron Radiation Research Institute, SPring-8, 1-1-1 Kouto, Sayo, Hyogo 679-5198, Japan}
\author{Y. Sakurai}
\affiliation{Japan Synchrotron Radiation Research Institute, SPring-8, 1-1-1 Kouto, Sayo, Hyogo 679-5198, Japan}

\date{\today}

\begin{abstract}
We report high-resolution, bulk Compton scattering measurements unveiling the Fermi
surface of an optimally-doped iron-arsenide superconductor,
Ba(Fe$_{0.93}$Co$_{0.07}$)$_2$As$_2$. Our measurements are in agreement with
first-principles calculations of the electronic structure, revealing both the
$X$-centered electron pockets and the $\Gamma$-centered hole pockets. Moreover,
our data are consistent with the strong three-dimensionality of one of these
sheets that has been predicted by electronic structure calculations at the local-density-approximation-minimum As position. Complementary calculations of the noninteracting susceptibility, $\chi_0({\bf
q}, \omega)$, suggest that the broad peak that develops due to interband Fermi-surface nesting, and which has motivated several theories of superconductivity
in this class of material, survives the measured three dimensionality of the Fermi surface in this family.
\end{abstract}

\pacs{71.18.+y, 74.70.-b, 78.70.Ck}

\maketitle 

The discovery of superconductivity in LaO$_{1-x}$F$_x$FeAs,
a member of the so-called ``1111'' family of iron-pnictide
materials \cite{kamihara2008}, has triggered intense research into its
nature and origin.  Tempted by the proximity of the superconductivity in
these compounds to an antiferromagnetic spin density wave (SDW) state,
many theoretical models have focused on spin fluctuations as the key
\cite{mazin2008a,kuroki2008}, supported by first-principles calculations
of the electronic structure. Such calculations predict four Fermi-surface (FS)
sheets: two hole sheets at the center of the Brillouin zone (BZ), and two
electron sheets centered at its corner \cite{singh2008a,singh2008b}. In
the undoped (and lightly doped) ``1111'' compounds the hole and electron sheets are
well nested, leading to a broad peak in the noninteracting susceptibility
\cite{mazin2008a}. The spin fluctuations that are thought to develop from this
FS instability have already been observed in neutron scattering measurements
\cite{neutron}, and theoretical models
suggest they promote a strong pairing interaction, provided that the order
parameter changes phase between the hole and electron FS sheets (the so-called
$s_{\pm}$-model) \cite{mazin2008a,kuroki2008}. Central to such models, however,
is a detailed understanding of the topology of the FS; here we present a bulk
measurement of the FS of an optimally-doped iron-arsenide superconductor,
Ba(Fe$_{0.93}$Co$_{0.07}$)$_2$As$_2$, alongside complementary calculations of the
noninteracting susceptibility, $\chi_0({\bf q}, \omega)$. We demonstrate the data are most 
consistent with a three-dimensional (3D) FS but that the nesting that is required for the $s_\pm$-model survives, lending its availability to all iron-pnictide compounds.

Ba(Fe$_{1-x}$Co$_x$)$_2$As$_2$ is a member of the ``122'' family of
iron-pnictide superconductors, with a maximum $T_c\sim24$\,K for $x\sim0.06$
\cite{canfield2008,chu2009}. Unlike the ``1111'' family, whose FS is quasi-two-dimensional (2D),
the ``122'' compounds are predicted by electronic structure calculations to
have substantial 3D warping of one of the hole sheets \cite{singh2008b},
the degree of which is sensitive to the position of the As with respect
to the Fe plane. Experimental studies have not yet adequately resolved wether 
the electronic structure is better described by the experimental As position or that at the 
local density approximation (LDA) minimum. Nonetheless, both families exhibit nesting between
parallel pieces of the hole and electron FS sheets.  Neutron scattering
measurements reveal spin fluctuations at the FS nesting vector, which is
coincident with the antiferromagnetic (AFM) wave vector for both ``1111'' and ``122'' compounds
\cite{neutron}.  For Fe(Te,Se), where the
nesting vector is not coincident with the AFM wave vector, spin fluctuations
are only evident at the FS nesting vector \cite{fetese},
indicating that the long-range magnetic order is not FS driven
\cite{johannes2009}.

Experimentally, de Haas-van Alphen (dHvA) measurements have found good
agreement with LDA calculations, observing orbits
that correspond to both electron and hole FS sheets in LaFePO \cite{coldea2008}
and non-superconducting SrFe$_2$P$_2$ \cite{analytis2009}.  Indeed, only
small rigid shifts of the LDA band structure were required in both cases
to achieve good agreement with the experimental data. However, owing to 
restrictions imposed by the electron mean free path, these measurements
have so far been confined to compounds away from the maximum Tc.
Angle-resolved photemission (ARPES)
measurements, on the other hand, yield a more contradictory picture. Whereas
some measurements on the undoped or lightly-doped BaFe$_2$As$_2$ system
demonstrate reasonable broad, albeit qualitative,
agreement with LDA predictions \cite{malaeb2009cm,yi2009,arpes}, others
suggest either the existence of hole-like ``blades'' in place of the quasi-2D
electron sheets expected at the corner of the BZ \cite{zabolotnyy2009a}
that are interpreted as indicative of additional order, or do not simultaneously
observe both hole and electron FS sheets. Indeed,
scanning-tunneling microscopy has been applied to the surface
layers of Ba(Fe$_{0.93}$Co$_{0.07}$)$_2$As$_2$, finding a complex array
of sample-dependent (specifically, cleavage temperature) ordered and
disordered structures \cite{massee2009cm}, which are not representative
of the bulk electronic structure.

\begin{figure}[t]
\includegraphics[angle=90,width=0.90\linewidth,clip]{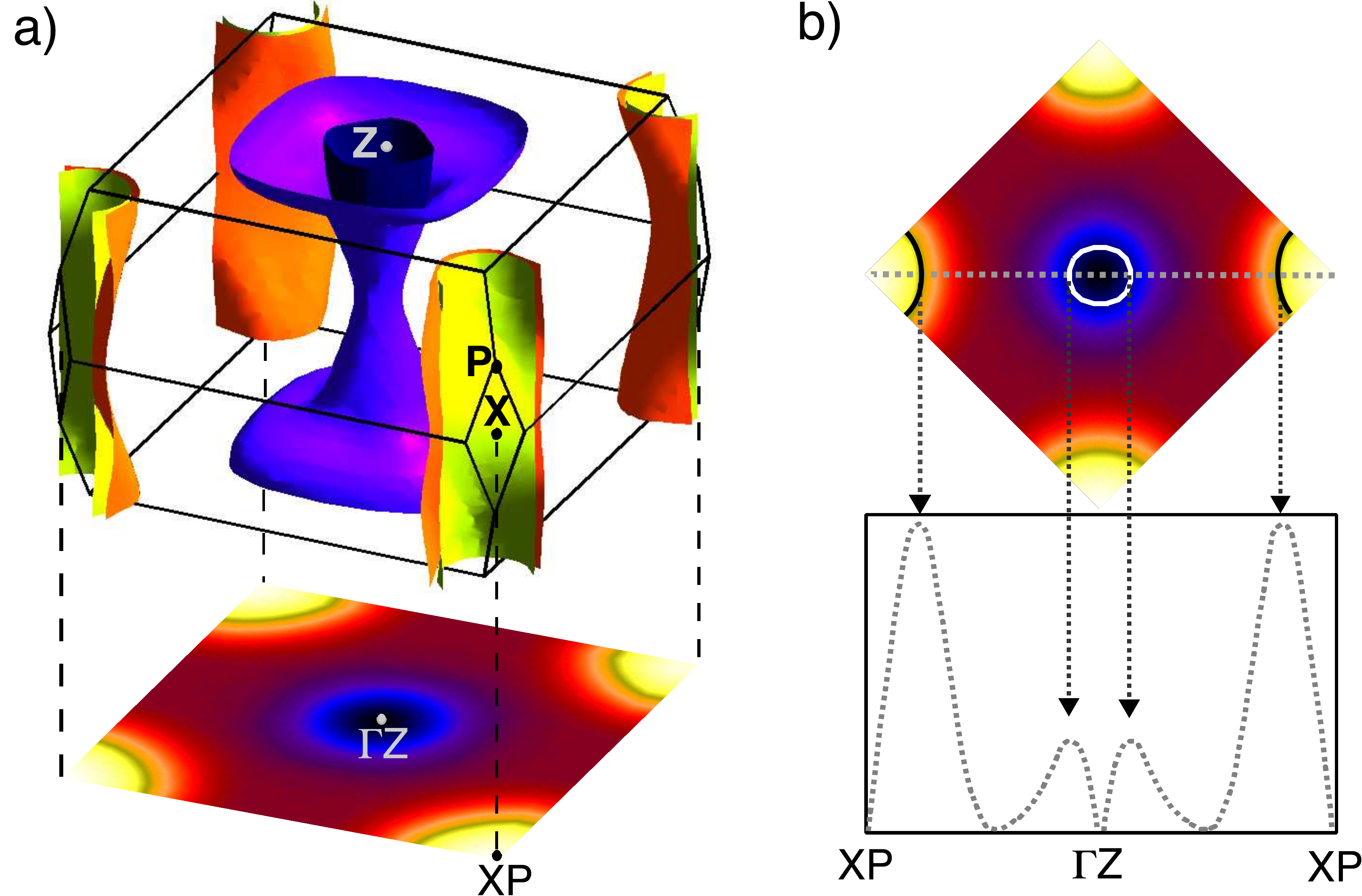}
\caption{\label{fig:fs} (Color online) (a) The LDA FS of
Ba(Fe$_{0.93}$Co$_{0.07}$)$_2$As$_2$ (top) and its relation to the occupation density
projected down the $c^*$ axis (bottom). Note that the occupation density
has been convoluted with the experimental resolution function.
(b) The overlaid FS contours (black and white lines, top) are identified via the
maxima in (the absolute value of) the first derivative of the occupation density shown at the 
bottom along a projected, high symmetry path.}
\end{figure}

More recently, several ARPES studies
\cite{vilmercati2009,malaeb2009cm,liu2009}, supported by dHvA
\cite{analytis2009} measurements, have reported on the 3D nature of the ``122'' family
FS, and particularly the warping of one of the hole FS sheets that has been
predicted by electronic structure calculations \cite{singh2008b}. The
FS that has motivated spin-fluctuation models of
superconductivity has so far been quite 2D, employing either the quasi-2D
band structure of LaFeAsO$_{1-x}$F$_x$ \cite{mazin2008a} or a 2D minimal
model of it \cite{kuroki2008}. Since the warping is expected to extend over some quarter
of the $c^*$ axis, the question of whether the nesting instabilities that
are expected to drive the spin fluctuations are robust against such three-dimensionality must be raised.

Compton scattering provides a valuable complement to dHvA and ARPES,
unambiguously probing the occupied states of the {\em bulk} electronic structure \cite{cooper2004}. Its
insensitivity to defects or disorder make it an ideal technique for tackling
the FS of disordered systems (e.g.\ \cite{dugdale2006}) and, since it is
sensitive to the bulk 3D momentum density, it is able to provide information
about the FS throughout the BZ.  In this study the occupation density,
featuring the projected Fermi breaks, serves as the key quantity for assessing
the FS of Ba(Fe$_{0.93}$Co$_{0.07}$)$_2$As$_2$.

\begin{figure}[t]
\includegraphics[width=0.90\linewidth,clip]{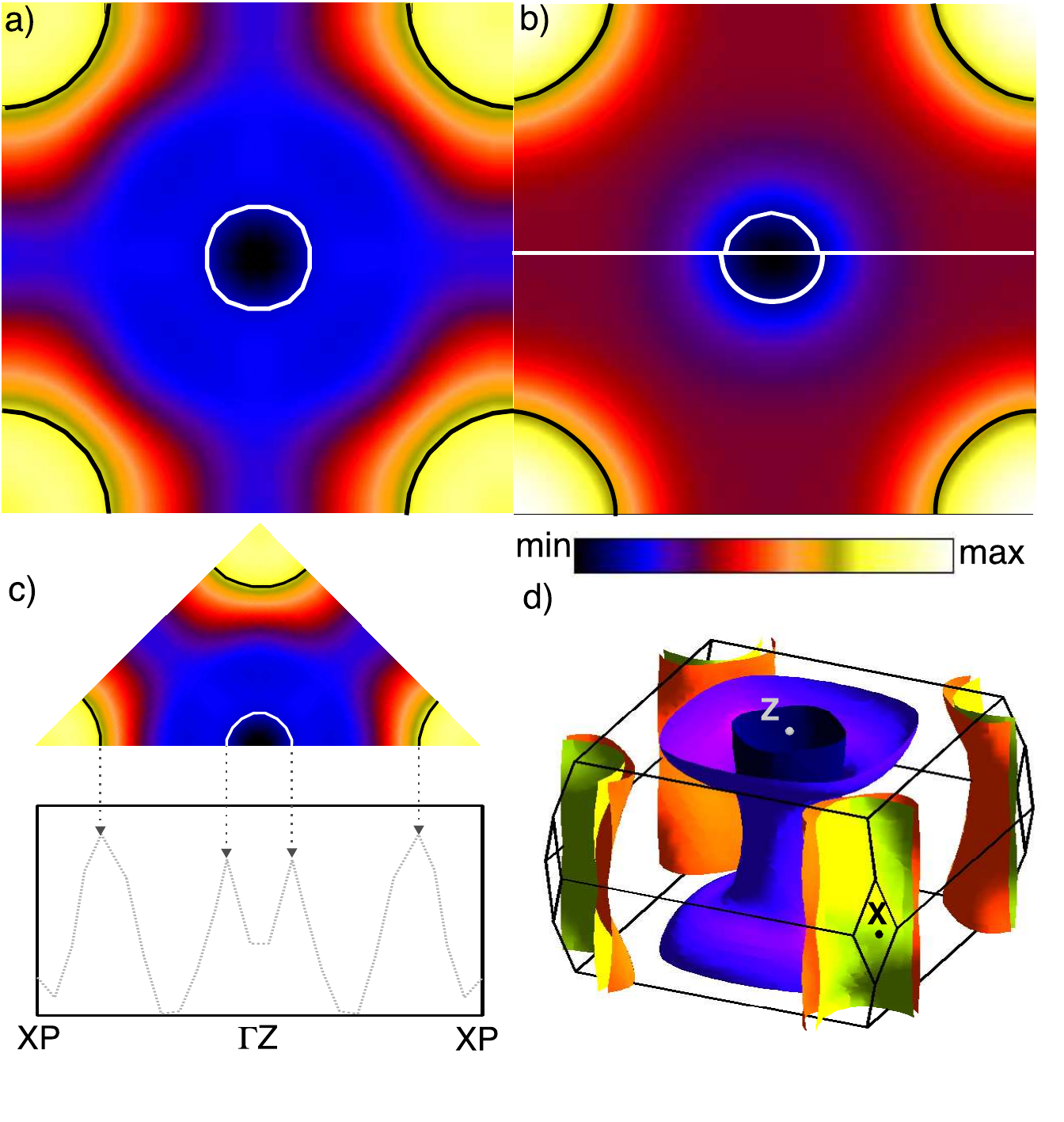}
\caption{\label{fig:occs} (Color online) The occupation density of
Ba(Fe$_{0.93}$Co$_{0.07}$)$_2$As$_2$ from (a) experimental Compton scattering
measurements at $295$\,K and (b) the electronic structure calculations of Fig.\
\ref{fig:fs}a from the original calculation (top), and after rigidly shifting
the bands (see text, bottom). The first derivative of the experimental data is shown in (c) in the same way as in Fig.\ \ref{fig:fs}b. In (d) the FS corresponding to the shifted
bands is shown for comparison with Fig.\ \ref{fig:fs}a.}
\end{figure}

A Compton profile represents a double integral [one-dimensional (1D) projection] of
the underlying 3D electron momentum density (EMD) in which the
FS manifests itself via discontinuous breaks.  Six Compton profiles
equally spaced between [100] and [110] were measured at room temperature ($T=295$\,K) on
the high-resolution Compton spectrometer of beamline BL08W at the SPring-8
synchrotron \cite{spring8}, with
a resolution full width at half maximum at the Compton peak of 0.105 a.u.($1/8$th of the BZ).  For each Compton profile,
$\sim 480\,000$ counts were accumulated in the peak data channel, and each
Compton profile was corrected for possible multiple-scattering contributions 
\cite{sakai1987}.  Details of the sample growth
and characterization may be found in Ref.\ \cite{chu2009}.

For the purpose of comparison with the occupation density
predicted by band theory and to aid interpretation of our data, we
performed {\it ab initio} calculations of the electronic structure of
Ba(Fe$_{0.93}$Co$_{0.07}$)$_2$As$_2$ using the full-potential linearized augmented
plane wave method
\cite{wien2k} within the LDA, where the effect of
doping was included via the virtual crystal approximation (VCA) \cite{paras}.
The As positions are those corresponding to the LDA minimum \cite{singh2008b}. 
Fig.\ \ref{fig:fs}a shows the calculated FS, composed of two
quasi-2D electron sheets and two hole sheets, in agreement with previous
calculations for similar Co concentrations \cite{singh2008b,sefat2008}. Note
the considerable three-dimensionality (``warping'') of the outer hole sheet.

Using tomographic techniques \cite{kontrym-sznajd1990}, it is possible to
reconstruct the 2D EMD from a specially
chosen set of 1D Compton profiles. Here, our six Compton profiles were
reconstructed to obtain the 2D EMD projected along the $c^*$ axis, which was
then folded back into the first BZ \cite{lock1973}. This culminates in a 
projected occupation density, shown for
the data in Fig.\ \ref{fig:occs}a, representing a projection of the 
occupied electron states in the BZ, where light (dark) colors represent high
(low) occupation densities. As can be seen from the data, both the electron
and hole FS sheets that are predicted by electronic structure calculations
(Fig.\ \ref{fig:occs}b) are clearly identifiable, with the electron sheets
in the corner of the projected BZ ($XP$) and the hole sheets in the center,
$\Gamma Z$. The slight flattening in the measured occupation density 
midway between $\Gamma Z$ and $XP$
arises as a consequence of describing its rapidly varying anisotropy with a finite 
number (six) of measured profiles and does not interfere with the location of the 
FS breaks. It is worth emphasizing that although this quantity is 2D,
information on the third, projected dimension is still stored in the integral.
For an ideally 2D FS, the Fermi breaks would be realised as a step function
between high and low occupation densities broadened only by the experimental
resolution; however, the three-dimensionality of the FS results in an additional smearing 
of these projected breaks.

In order to refine this picture, a method for
detecting the FS breaks via the first derivative of the occupation density,
as illustrated in Fig.\ \ref{fig:fs}b for the theoretical result, was employed.
The subtraction of an isotropic component, corresponding to the tightly-bound 
electron states, from the data before applying the folding procedure \cite{lock1973}
yields further enhancement in locating the Fermi breaks and hence the maxima
in the first derivative of this quantity were used to locate the breaks 
(see Fig.\ \ref{fig:occs}c), and the occupation
density was contoured at the corresponding level \cite{note}. Fig.\ \ref{fig:occs}a
shows overlaid on the data the experimental Fermi-surface breaks mapped out
utilizing this method. Owing to the experimental resolution and the projected
nature of this result, two features should be noted: first, the two electron
(hole) sheets are not resolved individually and second the contours here
represent an average over the shape and size of the FS sheets projected
down the $c^*$ axis. The slightly larger size of the electron compared with
the hole pockets is expected as a consequence of electron doping. For these
(experimental) contours assuming a quasi-2D FS topology, this leads to an 
occupied volume of the BZ of $4.38 \pm 0.16$ electrons. In comparison, 
for Ba(Fe$_{0.93}$Co$_{0.07}$)$_2$As$_2$,
the FS should enclose $4.14$ electrons, and we now focus on an explanation
for the discrepancy between these two figures.

In Fig.\ \ref{fig:occs}b, the same procedure that was applied to the data has been
employed for the theoretical FS, which explicitly includes the
three-dimensionality of the outer hole sheet, and good agreement is found
between experiment and theory. However, our method for locating the Fermi breaks
relies on a peak in the first derivative of the occupation density, and as
demonstrated in Fig.\ \ref{fig:fs}b, no such peak is discernible in the
theoretical occupation density that corresponds to the wider part of the
``warped'' hole FS. In projection, and with our experimental resolution, this
feature is not resolvable and is therefore expected to be absent in the
experimental contours as well. On the other hand, since some of the hole FS
is not accounted for in the contours, the ``warping'' still leaves its mark in the
occupation density in the form of an imbalance in the detected electron count.
Indeed, we find that $4.34$ electrons are enclosed by our theoretical contours,
where the excess electrons stem from an {\em underestimation} of the volume of
the hole FS, i.e., as a direct consequence of the ``warping'' of one of those
cylinders. Since the theoretical 3D FS used in this analysis unambiguously encloses
$4.14$ electrons, this supports well the notion that the extra electrons in our
experimental contours provide strong evidence of the 3D warping of the hole FS.
It should be noted that the theoretical FS is calculated with the As atoms at
the LDA minimum \cite{singh2008b}; crucially, our calculations at the experimental
As position, in which the ``warping'' is much weaker, were not able to explain such
a mismatch in the occupied fraction of the BZ.

Turning our attention to the details of the FS contours, and their
comparison with theory, our data suggest slightly larger FS sheets than
those expected from the calculation. It is instructive to investigate
whether such a discrepancy can be accounted for by a rigid shift of the
theoretical band energies, in such a way that the electron sheets increase
in size while the overall electron count in the system is maintained by
a shift on the hole sheets. As can be seen in Fig.\ \ref{fig:occs}b,
shifts of $\sim -120\,{\rm meV}$ and $\sim +80\,{\rm meV}$ for the electron and
hole bands, respectively, lead to a good overall agreement between data and
theory in which the shifted theoretical contours now enclose $4.40$ electrons. 
As illustrated in Fig.\ \ref{fig:occs}d employing such shifts preserves the 
FS topology, only affecting the size of the hole and electron pockets.

Overall, the projected FS topology obtained in our study is in good
agreement with other measurements on similar compounds (e.g. \cite{analytis2009, malaeb2009cm,yi2009}), as
well as electronic structure calculations within the LDA. In Ref.\
\cite{analytis2009}, a rigid shift of the theoretical electronic
structure was required to well reproduce their quantum oscillations
measurements on SrFe$_2$P$_2$. Those shifts were of a similar magnitude,
and were in the opposite direction; however, the necessary
additional approximations employed in this calculation, specifically the
VCA which is an average potential approximation may not provide a complete
picture of the disorder in contrast to a more sophisticated approach such as the 
coherent-potential approximation \cite{gyorffy72}.  Additionally, our data show 
no evidence of the propeller-like
structure near the $X$-point that has recently been reported in some
ARPES measurements \cite{zabolotnyy2009a}.

\begin{figure}[t]
\includegraphics[width=1.\linewidth,clip]{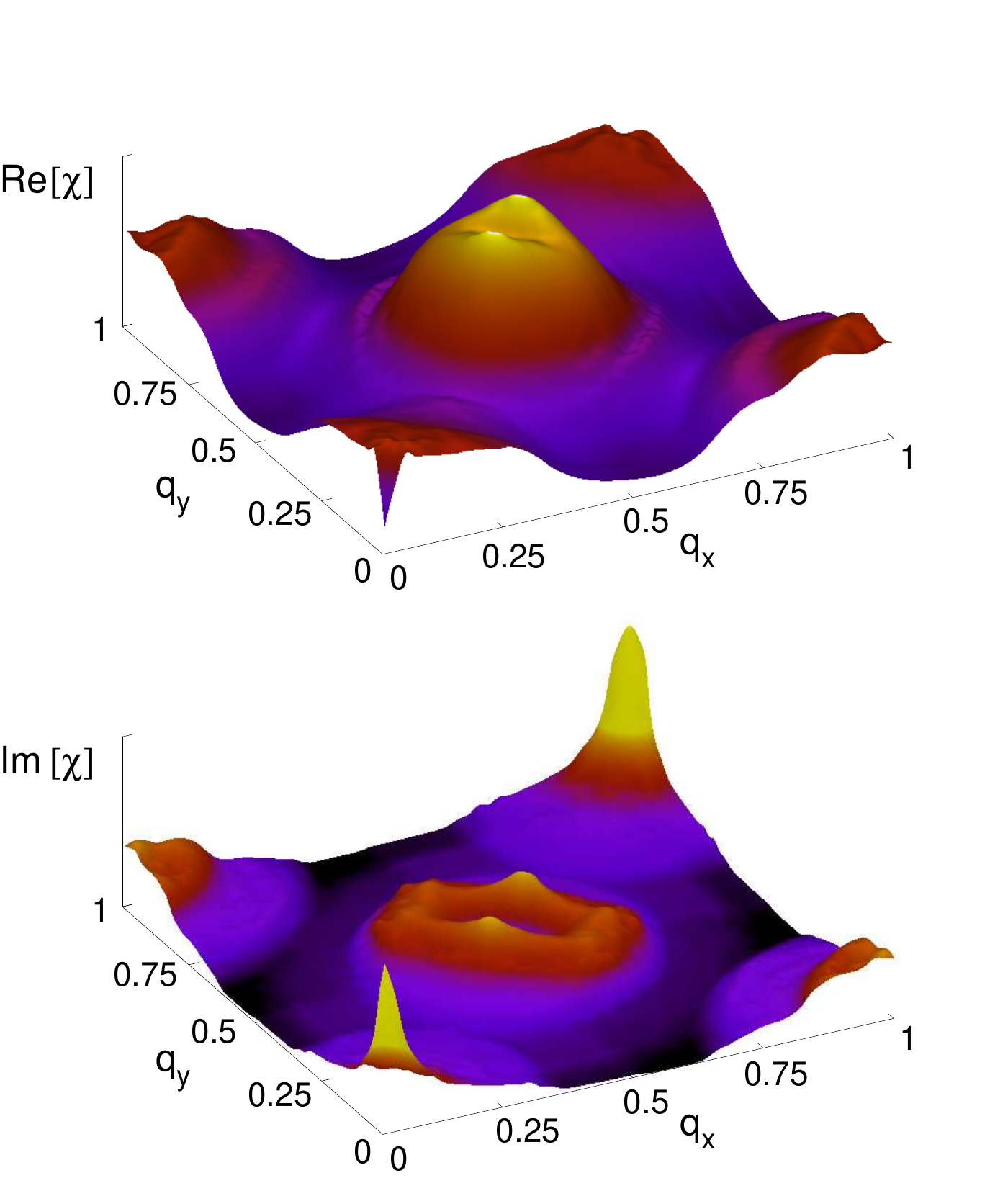}
\caption{\label{fig:chiq} (Color online) The real and imaginary
parts of the static susceptibility, $\chi({\bf q},0)$, of
Ba(Fe$_{0.93}$Co$_{0.07}$)$_2$As$_2$. The wave vector ${\bf q}$ is expressed in units of $2\pi/a$.}
\end{figure}

The nesting between these hole and electron FS sheets has been
implicated in several models of superconductivity, motivated by
a quasi-2D FS topology that provides ample nesting throughout the
$c^*$ axis of the BZ \cite{mazin2008a,kuroki2008}, as evidenced by
the broad peak in the susceptibility near ${\bf q} = (\pi,\pi)$
\cite{mazin2008a}.  However, for the ``112'' compounds presented
here, our data, in agreement with some ARPES and dHvA studies
\cite{vilmercati2009,malaeb2009cm,liu2009,analytis2009}, suggest that
an appreciable portion of the FS along the $c^*$ axis does not nest at
the same wave vector.  In order to address what impact this may have on the
susceptibility of the system, we have calculated the
noninteracting susceptibility, $\chi_0({\bf q}, \omega)$, for wave vector
${\bf q}$ and frequency $\omega$ \cite{mazin2008a}. A previous calculation of $\chi_0({\bf q}, \omega)$ of the ``122'' materials \cite{yaresko09} has
focussed on the stoichiometric and hole-doped compounds, and with the
As atom located at the experimental position, for which the
three-dimensionality of the Fermi surface is substantially less
pronounced. Both real and imaginary parts of
$\chi_0({\bf q}, \omega)$ were calculated for the (raw, unshifted) electronic structure
of Ba(Fe$_{0.93}$Co$_{0.07}$)$_2$As$_2$, and are shown in Fig.\ \ref{fig:chiq}.
Despite the three-dimensionality of the FS, a substantial (although broad)
peak in the imaginary part of $\chi_0({\bf q})$, whose origin lies in FS
nesting, survives into the real part, with a maximum at ${\bf q}_{\rm max}
\approx (0.56,0.56) \pi/a$. This broadly peaked structure bears close
resemblance to that calculated for the ``1111'' compounds in Ref.\
\cite{mazin2008a}, although the strength of the peak is slightly weaker here due
to the ``warping'' in $k_z$. The wave vector of the maximum itself, ${\bf q}_{\rm max}$
agrees favorably with the nesting
vector obtained from our experimental contours of ${\bf q}_{\rm exp} \approx
(0.62,0.62) \pi/a$, given that our contours represent an {\em average} of
the FS sheets [applying the same procedure to the convoluted theoretical
occupation density yields ${\bf q}_{\rm the} \approx (0.61,0.61) \pi/a$].
That our experimental results are in good agreement with the LDA calculation,
particularly with the topology of the FS, provides experimental evidence that
the $s_{\pm}$-model may still provide a route to superconductivity in the
``122'' compounds, despite the enhanced three-dimensionality of the electronic
structure.

In conclusion, the FS topology of Ba(Fe$_{0.93}$Co$_{0.07}$)$_2$As$_2$
has been investigated using Compton scattering, a true bulk probe of the
occupied electron states. Good agreement is found between our experimental
projected FS and the predictions of the LDA, which can be maximised by employing
only small rigid shifts of the theoretical electronic structure.
In particular, our data cannot be explained without including
the strong three-dimensionality of the hole FS expected for the ``122''
family at the LDA-minimum As position (rather than the experimental one). 
However, our calculations of the noninteracting susceptibility
of this compound suggest that the peak that has been implicated in some
theories of superconductivity survives this three-dimensionality and may
still be available as a mechanism in these compounds. Such an observation 
provides experimental evidence that the $s_\pm$-model (and others that rely 
on FS nesting) may be general models for superconductivity in the iron pnictides. 

\section*{Acknowledgments}
We would like to thank C.\ Lester and S.M.\ Hayden for their technical assistance. We
acknowledge the financial support from the UK EPSRC-GB. This experiment was
performed with the approval of the Japan Synchrotron Radiation Research
Institute (JASRI, proposal no.\ 2009A1094). Work at Stanford University is supported 
by the Department of Energy, Office of Basic Energy Sciences under Contract No. 
DE-AC02-76SF00515.


\begin{thebibliography}{99}
\bibitem{note}
Applying the same procedure to our 
calculations was found to yield a relative enhancement of the
projected Fermi breaks, without influencing the location at which they
were identified.
\end{thebibliography}
\end{document}